\documentclass[aps,pre,superscriptaddress,showpacs,reprint]{revtex4-1}
\usepackage[usenames,dvipsnames]{xcolor}
\usepackage{times}
\usepackage{graphicx}
\usepackage{amssymb}
\usepackage[fleqn]{amsmath}
\usepackage{epsf}
\usepackage{hyperref}
\usepackage{color}
\usepackage{verbatim}
\usepackage{ulem}

\newcommand{\ket}[1]{\left| #1 \right>} 
\newcommand{\bra}[1]{\left< #1 \right|} 

\begin{document}
\title{Environment-induced anisotropy and sensitivity of the radical pair mechanism in the avian compass}
\author{Alejandro Carrillo}
\affiliation
{Instituto de F\'isica Gleb Wataghin, Universidade Estadual de Campinas, CEP 13083-859, Campinas, S\~ao Paulo, Brazil}
\author{Marcio F. Cornelio}
\affiliation{Instituto de F\'isica, Universidade Federal de Mato Grosso, 78068-900, Cuiab\'a MT, Brazil}
\author{Marcos C. de Oliveira}
\affiliation
{Instituto de F\'isica Gleb Wataghin, Universidade Estadual de Campinas, CEP 13083-859, Campinas, S\~ao Paulo, Brazil}

\begin{abstract}
{Several experiments over the years have shown that the Earth's magnetic field is essential for orientation in birds migration. The most promising explanation for this orientation is the photo-stimulated radical pair (RP) mechanism. 
In order to define a reference frame for the orientation task radicals must have an intrinsic anisotropy. 
We show that this kind of anisotropy, and consequently the entanglement in the model, are not necessary for the proper functioning of the compass.
Classically correlated initial conditions for the RP, subjected to a fast decoherence process, are able to provide the anisotropy required. 
Even a \textit{dephasing} environment can provide the necessary frame for the compass to work, and also implies fast decay of any quantum correlation in the system without damaging the orientation ability. This fact significantly expands the range of applicability of the RP mechanism providing more elements for experimental search.}
\end{abstract}
\maketitle




\section{Introduction}

The ability of birds to use the Earth's magnetic field to orientate themselves in the correct direction for migration \cite{Wiltschko1,Wiltschko2} has originated several experimental works devoted to the understanding of the main features of the underlying mechanism \cite{Wiltschko5,Wiltschko3, Wiltschko4,Wiltschko9}.
One of the first proposals for modeling the appearance of the magnetic compass was that magneto-perception operates by means of anisotropic magnetic field effects on the rate of production of yields of a photo-stimulated radical pair reaction \cite{Schulten1,werner,SchultenBook}.
Although other models have been proposed, 
such as a magnetite-based magneto-perception \cite{Kirschvink, Johnsen}, the proposal of a radical pair mechanism (RPM) has been recently reinforced \cite{Ritz3} and strong {experimental} and theoretical evidences have been presented in its favor \cite{Ritz1,Cintolesi,Timmel,Wiltschko10,Natcom,Maeda}; suitable molecular candidates for mechanism are cryptochrome photoreceptors \cite{Wiltschko10,Ritz2011}.
The solid state RPM model can be summarized as follows \cite{Ritz3}: a molecular precursor reacts to form a pair of radicals due to photochemically-driven electron transfer.
Taking into account that both radicals are created in a single event, it is natural to assume that the electron spins are initially entangled (in the following we assume that the radical pair (RP) is created in a singlet $s$ state, although working with a triplet $t$ state is also possible). This singlet state evolves under the influence of a Hamiltonian containing {an} hyperfine interaction term between the nuclei and their electrons and a Zeeman interaction term between the unpaired electrons. Due to the anisotropy of the hyperfine tensor \cite{Boxer1},
the interconversion between entangled singlet and triplet states depends on the direction of the applied magnetic field through the Zeeman term in the Hamiltonian.
It is necessary also to assume that the radicals are almost immobile, without significant diffusive motion, in order to avoid the anisotropy present in the system to be averaged away.
The RP yields depend on the relative alignment of the magnetic field in relation to the sample \cite{Boxer1,Boxer2}, so that it can work as a compass.

In this work, we show {that} the anisotropy in the molecule can be replaced by an anisotropic environment. This fact allows for a free isotropic molecule in a diffusive environment to work as a compass. {The model is derived through the inclusion of dipole-dipole interaction, which although being weak for each individual pair, can effectively account for the required anisotropy.}
In addition, although some discussion have been given recently about the importance of entanglement in the magneto-{perception} process \cite{Cai, Gauger}, this still remains obscure. In this sense, we also find that entanglement is neither necessary in our model of isotropic molecules nor in the anisotropic ones.
Furthermore, we also verify the functioning of the compass in the presence of artificial radio frequency fields. We find that the isotropic model cannot work in the presence of such field in agreement with experimental findings \cite{Ritz2,Ritz1,Wiltschko9}. However, the anisotropic model can work under some environmental conditions not unlikely in an open system, which notwithstanding disagree with experimental observation. This fact gives more one evidence in favor of the present model of an isotropic molecule {together with the environment induced anisotropy}. 
%

We begin by reviewing in Sec. \ref{model} the basic model for the avian compass based on photo-stimulated radical pair reaction \cite{Gauger}. In Sec. \ref{noise} we describe the models employed for environmental noise. In Sec. \ref{anisotropy} we discuss our main results regarding the environmental induction of anisotropy and finally in Sec. \ref{disc} we conclude the paper.

\section{Model}\label{model}
 
Let us consider the Hamiltonian of the RP, neglecting {for the moment} all other possible interactions, such as exchange and dipolar, to be:
\begin{equation}\label{eq:hamiltonian}
\hat{H}^{k}=\sum_{i}\mathbf{\hat{I}}_{ik}\cdot A_{ik} \cdot \mathbf{\hat{S}}_{k} + \omega_{e}\mathbf{B}\cdot \mathbf{\hat{S}}_{k},
\end{equation}
where the first term is the hyperfine contribution and the second one is the Zeeman contribution, with $i$ labeling the $i$-th nucleus in the $k$-th radical. Earth's magnetic field is given by $\mathbf{B}=B_{0}(\sin\theta \cos\phi,\sin\theta \sin\phi, \cos\theta)$, and $A$ is the hyperfine tensor. $\mathbf{\hat{I}}$ and $\mathbf{\hat{S}}$ are the spin operators for the nucleus and the electron respectively, $\omega_{e}=g_{e}\mu_{B}/\hbar$, $g_{e}$ is the electron g-factor, $B_{0}=47\mu T$ and $\mu_{B}$ is the Bohr Magneton. We assume that the electron $g_{e}$-tensor is isotropic and close to that of a free electron, and that the hyperfine tensor $A$ is, for simplicity, 
diagonal. Additionally, it can be either isotropic or anisotropic according to the model.
The direction of the applied magnetic field 
in relation to the RP fixed axis system is defined in terms of the polar angles $(\theta,\phi)$; without losing generality we are going to assume $\phi=0$ in order to simplify procedures. The most appropriate units to work with are those of Nuclear Magnetic Resonance (angular frequency units).

In order to fully determine the dynamics of our system, we are going to use a master equation approach \cite{Gauger}. 
To account for the singlet or the triplet yields, we add their formation rate 
to the master equation as a dissipative process modulated by a factor $k$ in angular frequency units. The simplest model we can consider is one nucleus coupled with its unpaired electron plus a free electron, {with Hamiltonian $\hat{H}=\mathbf{\hat{I}_1}\cdot A \cdot \mathbf{\hat{S}_1} +\omega_{e}\mathbf{B}\cdot \big(\mathbf{\hat{S}_1}+\mathbf{\hat{S}_2}\big)$}.
The evolution of the system can be written as:
\begin{align}\label{eq:MasterEquation}
\frac{d}{dt}&\rho(t)= i\big[\rho(t),\hat{H}\big]/\hbar\\ \notag
&+\sum_{n} \frac{k}{2} \big ( 2 P_{n} \rho(t) P_{n}-P_{n}\rho(t)-\rho(t)P_{n}\big )\\ \notag
&+\frac{\gamma}{2} \big( 2\hat{B}\rho(t) \hat{B}^{\dagger} -\hat{B}^{\dagger}\hat{B}\rho(t)-\rho(t) \hat{B}^{\dagger}\hat{B}\big). \notag
\end{align}
The operators $P_{n}$ are projectors over the singlet or triplet shells as described in \cite{Gauger}{; as with any projector, $P_{n}^{\dagger}=P_{n}$ and $P_{n}^{\dagger}P_{n}=P_{n}^{2}=P_{n}$}. {The sum is taken} over all possible {electronic and nuclear} spin states, {i.e. $\ket{l,m} \otimes \ket{\alpha}$, where $l$ is the electronic singlet or triplet state and $m$ its projection, and $\alpha$ represents the nuclear spin state, either $\uparrow$ or $\downarrow$.} 
The process mediated by $k$ can be thought as a measurement on the state of the RP, giving us information about the amount of singlet or triplet yield. However the density operator formalism enable us to take into account the environment in which it lives. A wide range of noise processes {tested to model environmental interactions resulted} in decoherence times of hundreds of microseconds \cite{Gauger,Bandyopadhyay2012}, although there has been some controversy \cite{Gauger2013} on the numerical results found in \cite{Bandyopadhyay2012}. 
 {The noise affecting the system can be either} \textit{amplitude} damping or \textit{dephasing} noise \cite{Marquardt2008}.
 
\section{Environmental noise} \label{noise}
\subsection{Amplitude Damping noise} 
 
\begin{figure}
  \centering
  \includegraphics[width=0.45\textwidth]{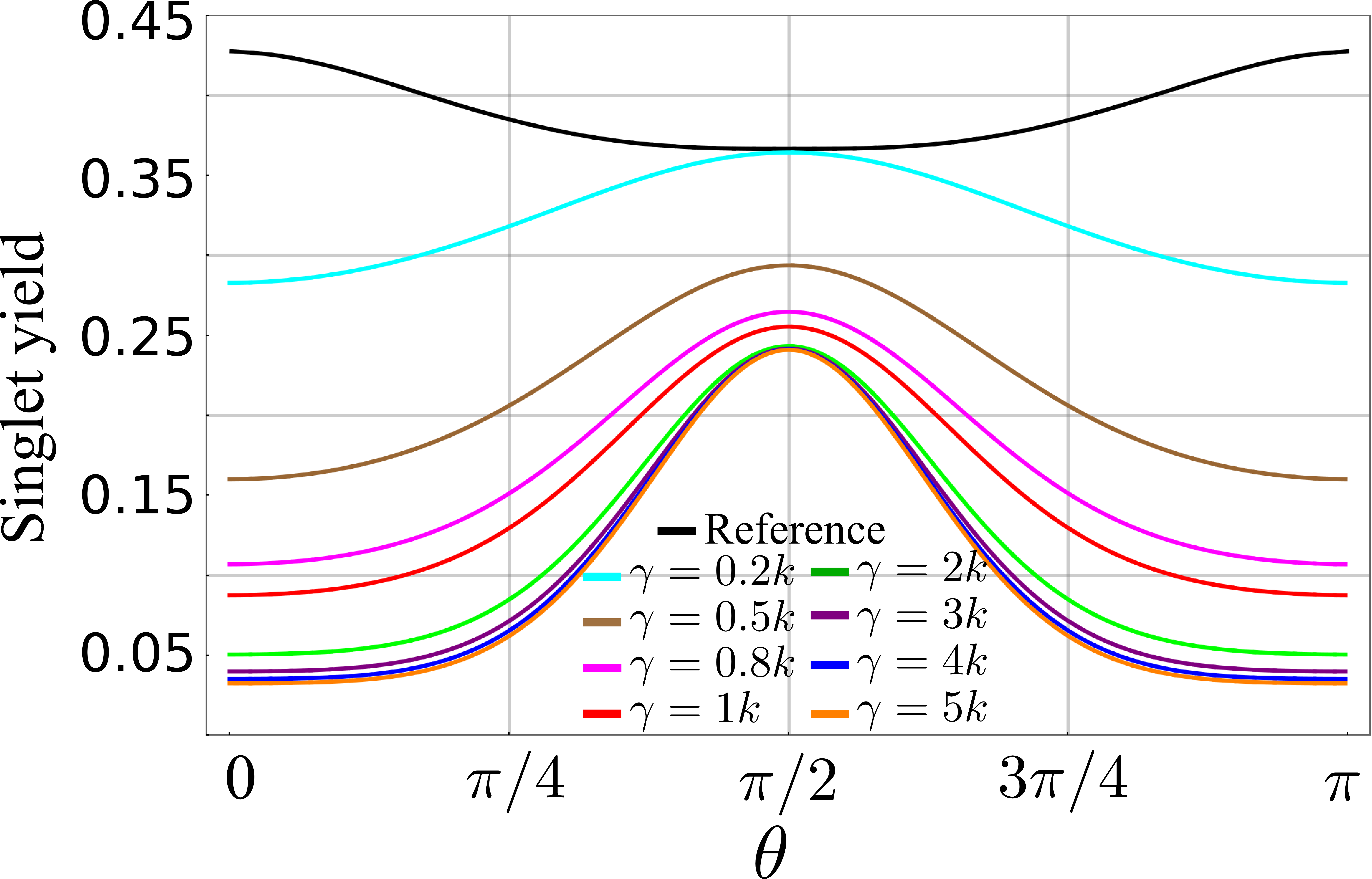}
  \caption{(Color online) Angle sensitivity of the radical pair with an isotropic hyperfine tensor $A=a\mathbb{I}$ in the presence of the amplitude damping noise described by the Lindblad operator \eqref{eqSM:lindblad} for different increasing values of the coupling constant $\gamma$  from top to bottom. The first curve at the top is a reference curve for $\gamma=0$, $k=0.01MHz$, for the anysotropic hyperfine tensor and initial singlet state.\label{figSM:damping}}
 \end{figure}

 The main characteristic of amplitude damping noise is that the operators that originate it do not commute with the Hamiltonian of the system; this has several implications on the way the channel affects its dynamics, being the most obvious that the energy is not preserved during its evolution. Moreover, amplitude damping introduces a somewhat obvious anisotropy in the system by promoting the interconversion between singlet and triplet states, and imposing non-uniform decay rates on the state coherences (non-diagonal terms).  
 As an example we use the operator $\sigma_{2}^{\dagger}$, which describes absorption of energy for the unpaired electron, raising its spin. Acting on the singlet state $\bra{S}$ it generates a triplet state with angular projection $m=1$, i.e.:
\begin{eqnarray}\label{eqSP:sigmap}
 \sigma_{2}^{\dagger}\ket{S}&=&\sqrt{2}\ket{T,1}.
\end{eqnarray}
This operator does not conserve the energy of the system. When used as a perturbative noise, it produces a Lindblad like term in the density matrix of the form:
\begin{eqnarray}\label{eqSM:lindblad}
 \mathcal{L}\rho &=& \frac{\gamma}{2}\big( 2 \sigma_{2}^{\dagger}\rho\sigma_{2}-\sigma_{2}\sigma_{2}^{\dagger}\rho-\rho\sigma_{2}\sigma_{2}^{\dagger}\big).
\end{eqnarray}
As a result of its action, the populations of the density operator will have an additional term of the form $\dot{\rho}_{\uparrow\uparrow}=-\gamma\rho_{\uparrow\uparrow}$ and $\dot{\rho}_{\downarrow\downarrow}=\gamma\rho_{\uparrow\uparrow}$. In Figure \eqref{figSM:damping} we show the behavior of the singlet yield when the system is in the presence of the noise process described above. It is evident that, as in the case of the dephasing noise in the main text, the damping process described here suffices to generate an angular sensitivity in the radical pair, although this should not be a surprise: the action of the operator on the states \eqref{eqSP:sigmap} of the system makes clear that the interconversion process is anisotropic, given that $\sigma_{2}^{\dagger}\ket{T,1}=0$. 
 \subsection{Pure Dephasing noise} 
 
 Our purpose is in the study of pure dephasing channels as a cause of angular sensitivity. This kind of channel would not affect the interconversion rate between singlet and triplet yields and in consequence is not imposing explicitly non-uniform decay rates. The operators originating pure dephasing noise commute with the Hamiltonian, giving rise to effects only on the coherences  of the density matrix. This channel could be originated due to dipole interactions, magnetic fluctuations in the biological environment \cite{Gauger} or even energy transfer through spin hoping \cite{Plenio2008} among others. 
Dipolar and exchange interactions between the electron pair in the radical have little influence in the angular sensitivity \citep{Efimova}. However a net effect of the influence of the molecules surrounding the radical pair can be a source of noise. From the dipolar interaction alone, three dephasing noise terms can emerge, although here we are going to focus our analysis in only one of them. The dipolar interaction can be written as
\begin{eqnarray}\label{eq:dipolar}
 H_{dip}&=&\frac{\mu_{B}}{4\pi r^5}\Big[3\big(\mathbf{m}_{1} \cdot \mathbf{r}\big)\big(\mathbf{m}_{2} \cdot \mathbf{r}\big)-r^2\mathbf{m}_{1}\cdot\mathbf{m}_2\Big] \\\notag
 &=&\frac{\mu_{B}g_{1}g_{2}}{8\pi r^3}\Big[3\big(\mathbf{\sigma}_{1} \cdot \hat{\mathbf{r}}\big)\big(\mathbf{\sigma}_{2} \cdot \hat{\mathbf{r}}\big)-\mathbf{\sigma}_{1}\cdot\mathbf{\sigma}_2\Big]\notag,
\end{eqnarray}
where the $g_{i}$ are the electronic factors of each electron, $\mathbf{\sigma}=\hat{e}_{x}\hat{\sigma}_x +\hat{e}_{y}\hat{\sigma}_y+\hat{e}_{z}\hat{\sigma}_z$ and $\hat{\mathbf{r}}$ is the unitary normal vector in spherical coordinates. If we consider $N$ environmental electronic spins interacting through eq. \eqref{eq:dipolar} with the unpaired electron in the radical pair, we will have a net dipolar contribution:
\begin{eqnarray}\label{eq:dipolarN}
 H_{dip}&=&\sum_{i}\frac{\mu_{B}g_{1}g_{2}}{8\pi r_{i}^3}\Big[3\big(\mathbf{\sigma}_{1} \cdot \hat{\mathbf{r}}\big)\big(\mathbf{\sigma}_{i} \cdot \hat{\mathbf{r}}\big)-\mathbf{\sigma}_{1}\cdot\mathbf{\sigma}_i\Big].
\end{eqnarray}
Before proceeding any further, we must note that due to the inter-molecular distances, the magnitude of the rate governing the interaction is going to be small compared with the production rate of the radical products $k$ described in the main text, and proportional to $1/r^3$. We are interested in showing how from this interaction a dephasing noise process may be obtained. To do that let us consider the Lindblad-like part of the master equation in the interaction picture:
\begin{eqnarray}\label{eq:MasterEquationSupl}
\frac{d}{dt}\rho(t)&=&-\int_{0}^{\infty}tr_{B}\Big\{ \big[H_{dip}(t),\big[H_{dip}(t-s),\rho(t)\big]\big]\Big\}ds,\nonumber\\
\end{eqnarray}
where $tr_{B}$ indicates the tracing of all the bath degrees of freedom. If the correlations of the system go to zero much faster than the natural time of the system $\tau_S$, we can write $\rho(t)=\rho_{S}(t)\otimes\rho_{B}$, where $\rho_{B}$ and $\rho_{S}(t)$ are the density matrix of the bath and the system respectively. For simplicity in the notation let's write $\hat{A}_{i}={\sigma}_{i} \cdot \hat{\mathbf{r}}$ and $\hat{A}={\sigma}_{1} \cdot \hat{\mathbf{r}}$. One of the terms emerging from the double commutator is:
\begin{eqnarray}
\big[H_{dip},\big[H_{dip},\rho_{S}(t)\otimes\rho_{B}\big]\big]&=&
\hat{A}^2\rho_{S}(t)\hat{A}_{i}\hat{A}_{j}\rho_{B}\nonumber\\&&+\rho_{S}(t)\hat{A}^2\rho_{B}\hat{A}_{i}\hat{A}_{j}\nonumber\\&&-2\hat{A}\rho_{S}(t)\hat{A}\hat{A}_{i}\rho_{B}\hat{A}_{j}.
\end{eqnarray}
Due to the cyclic permutation of the trace,  $tr_{B}\Big\{\hat{A}_{i}\hat{A}_{j}\rho_{B}\Big\}=tr_{B}\Big\{\rho_{B}\hat{A}_{i}\hat{A}_{j}\Big\}=tr_{B}\Big\{\hat{A}_{i}\rho_{B}\hat{A}_{j}\Big\}=tr_{B}\Big\{\rho_{B}\Big\}=1$. The remaining term of the double commutator after taking the trace is then:
\begin{align}\label{eq:MasterEquationSupl2}
\hat{A}^2\rho_{S}(t)&+\rho_{S}(t)\hat{A}^2-2\hat{A}\rho_{S}(t)\hat{A}= \\\notag
&-2\Big[2\big(1-2\hat{A}\big)\rho_{S}(t)\big(1-2\hat{A}\big)\\\notag
&- \big(1-2\hat{A}\big)^2\rho_{S}(t)-\rho_{S}(t)\big(1-2\hat{A}\big)^2\Big].\notag
\end{align}
This means that the dipolar interaction between the subsystems of the radical pair and the environmental degrees of freedom is a generator of dephasing channels. The procedures are similar for the remaining five operators that commute with the radical pair Hamiltonian, giving raise to five more dephasing channels. 
 {Finally then we introduce the dephasing channel in the Lindblad form in the last term of eq. \eqref{eq:MasterEquation}, where the operator $\hat{B}$ commutes with the Hamiltonian \eqref{eq:hamiltonian}; from \eqref{eq:MasterEquationSupl2} we have that $\hat{B}=\big(\mathbf{\hat{1}}-2 \sigma_{1} \cdot \mathbf{\hat{n}}\big)/\sqrt{2}$, where $\sigma_{i}= \hat{e}_{x}\hat{\sigma}_x +\hat{e}_{y}\hat{\sigma}_y+\hat{e}_{z}\hat{\sigma}_z$ and $\hat{\mathbf{n}}$ is the unitary normal vector in spherical coordinates. The rate of the process is given by $\gamma$, which represents a measure of the strength of the interaction between the system and the environment.}
 {All results presented in this paper were obtained by direct numerical integration of eq. \eqref{eq:MasterEquation}.}

\begin{figure}
 \centering
 \includegraphics[width=0.45\textwidth]{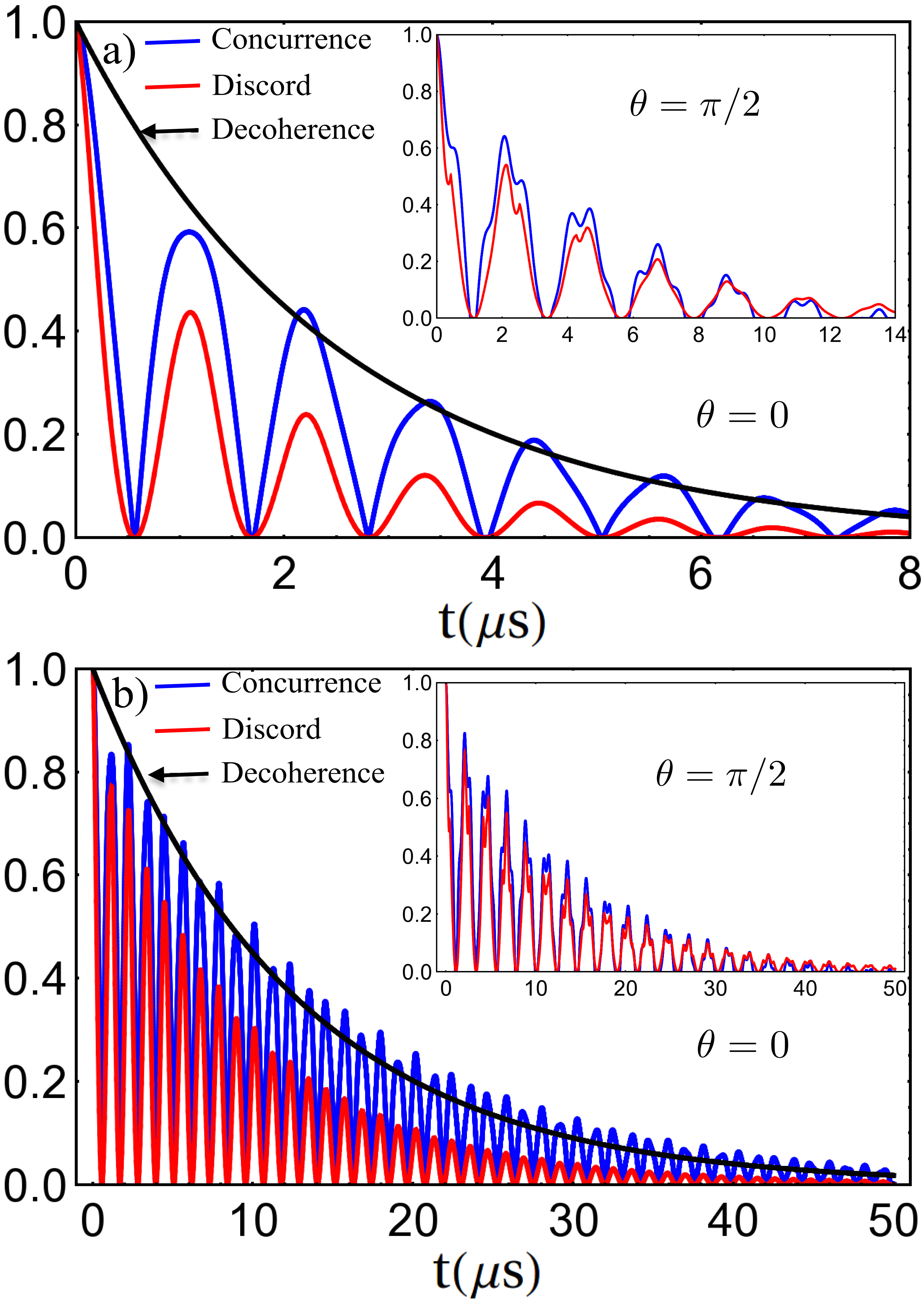}
 \caption{(Color online) Concurrence (top oscillating plot) and Quantum Discord (lower oscillating plot) evolution for different values of the rates and magnetic field inclination angles. a) Measurement rate $k=0.1MHz$, and dissipation rates $\gamma=2k$, with and angle $\theta=0$; in the inset an angle of $\theta=\pi/2$ was used. b) Measurement rate $k=0.01MHz$, and dissipation rates $\gamma=10k$ with and angle $\theta=0$; in the inset an angle of $\theta=\pi/2$ was used. Is worth to note that the angle affect the amount of quantum correlations measured by the QD and the Concurrence; for smaller angles the Concurrence values are not only higher, but predict the same time of decoherence as the QD; however, for higher angles, the entanglement is less important than the classical correlations, making the QD to be higher and to predict longer correlation times. The decoherence was computed by means of the Fidelity \citep{Nielse2011book} between the evolving state and the initial singlet state.\label{fig:CyQD}}
 \end{figure} 
 
\section{Anisotropies and angular sensitivity} \label{anisotropy}
The yields \textit{measurement} process, regardless of its magnitude, is not going to affect the performance of the compass. Therefore $k$ could be arbitrarily high and the compass would still work. However, an upper bound to $k$ can be determined by an important experimental observation:  It was observed, in a set of experiments with European Robins, that an oscillating \textit{rf} magnetic field $\vec{B}_{rf}$, perpendicular to the Earth's one, disrupts the avian compass functioning \cite{Ritz2,Ritz1}, leaving them without sense of direction. So when the magnitude of $k$ exceeds a threshold, the influence of the measurement process should overwhelm the action of the \textit{rf} field \cite{Gauger}, and as a result, the sensitivity to the variations of the magnetic field of Earth would be present despite of the disrupting effect. This fact can be contrasted to the experimental results to pick a suitable upper bound to $k$ as $k=0.01$MHz, which will be used unless stated otherwise. The 
processes mediated by $\gamma$ show the same behavior, i.e., their influence will allow the compass to work in spite of the presence of the \textit{rf} field if their magnitudes are high enough. As the experimental observation tell us that there will not be a compass if the birds are subject to this \textit{rf} field, we use this to set upper bounds for the noise amplitudes in the same order of magnitude of $k$. This fact also implies a lower limit for the decoherence time of the system, because an upper bound to the noise amplitudes implies that this time cannot be arbitrarily small, and we are going to have at least tens of microseconds until the loss of all the coherences in the system. We use a magnitude $B_{rf}=150$nT as proposed in \cite{Gauger}.

\subsection{Hyperfine tensor}
\begin{figure}
  \centering
  \includegraphics[width=0.45\textwidth]{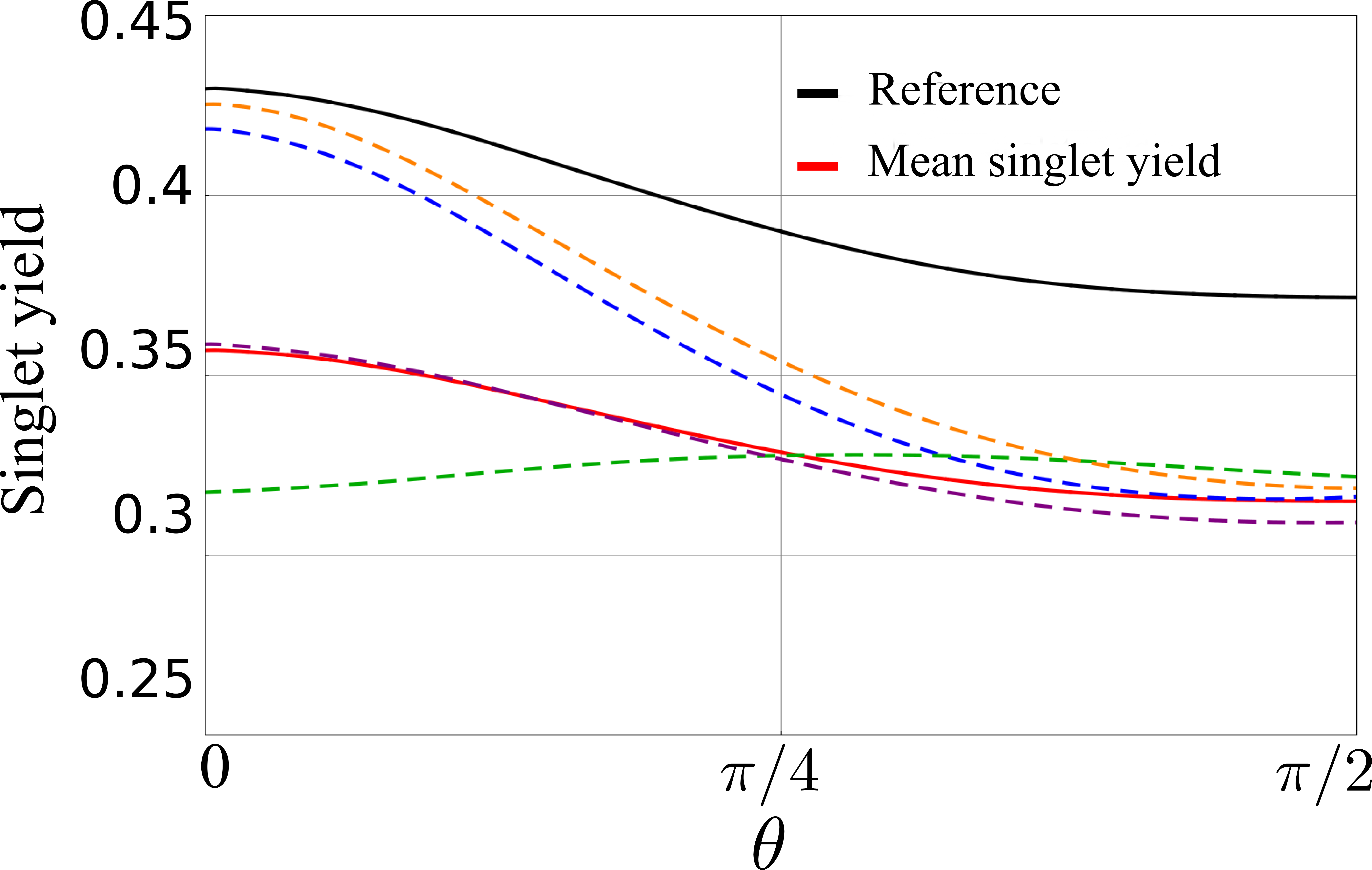}
  \caption{(Color online) Perturbations in a singlet initial condition with an anisotropic hyperfine tensor without environmental noise. The red (lower continuous) curve represents  the mean of $1000$ perturbed singlet initial conditions. The mechanism appears quite robust, as the differences in the yield
production remain unchanged when compared with the unperturbed singlet initial condition, viven by the black (top continuous) curve in the Figure). The dashed lines are four random yields taken from the sample.\label{figSM:random}}
 \end{figure}

As stated by Schulten \cite{Schulten1}, for the compass to work some kind of anisotropy must be present in the system. If, as usual, we choose the source of anisotropy {in} the hyperfine tensor, there is a sensitivity in the RPM to all initial states. To show that, we start by observing that for some values of the rates, for example $\gamma=2k$, the influence of the rf field over the compass is strengthened; and if $\gamma$ is of the order of $20k$ the compass-disrupting effect by the rf field can still be observed. It is interesting to note that even if in this situation the environment contributes to the insensibility of the compass by means of the rf field, in other circumstances can increase it \cite{PhysRevA.85.040304}. One of the consequences of using these processes is that the decoherence times are short. To give a measure of those times for testing the model with anisotropy in the hyperfine tensor we use the Quantum Discord (QD) \cite{Discord}, which is able to signal the 
presence of any kind of quantum correlations, and the Concurrence \cite{Wooters}, which measures quantum correlation as entanglement only. The former is defined as
\begin{equation*}
 \delta_{\overleftarrow{AB}}= I_{AB}-J_{\overleftarrow{AB}},
\end{equation*}
where $I_{AB}$ is the mutual information between $A$ and $B$ and is defined as $I_{AB}=S_{A}+S_{B}-S_{AB}$, $S_{x}$ is the von Neumann entropy of system $x$, $J_{\overleftarrow{AB}}=\max_{\{\Pi_{x}^{B}\}}\big[S(\rho_{A})-\sum_{x}p_{x}S(\rho_{A}^{x}) \big]$ is the classical correlation between the subsystems, $p_{x}=Tr_{A}\{\Pi_{x}^{B}\rho_{AB}\Pi_{x}^{B}\}$ and $\rho_{A}^{x}=Tr_{B}\{\Pi_{x}^{B}\rho_{AB}\Pi_{x}^{B}\}/p_{x}$; the maximum is taken over the positive measurements $\{\Pi_{x}^{B}\}$ made over the system $B$. The evolution of both, QD and Concurrence can be seen in Figure \ref{fig:CyQD}. {The reason for employing the QD, is that it signals the presence of quantum correlations even when there is no entanglement \cite{Ollivier2001}. Therefore if there is any quantum correlation preserving over long times, which could then be relevant, it would be signaled by QD.
However what both measures show is quite the opposite, i.e., they show that in general there is a rapid loss of any form of coherence compared to the results in \citep{Gauger}. The fast loss of coherence is not a surprise - having an open environment like the one we can expect in the eye of the bird, should lead naturally to a fast loss of quantum correlations. This result should not compromise the correct behavior of the compass; one of the reasons, concerning the role of entanglement in the system, will be presented below. It is also interesting to note that the change in the inclination of the magnetic field also affects the loss of coherence. This can be seen in Figure \ref{fig:ConcurrenceDensity}, where the change of concurrence with time and inclination angle, for $\gamma$ fixed, is shown. For small angles $\theta$ there are going to be more collapses and revivals of entanglement, and for angles near $\pi/4$, the decoherence time are shorter that for angles near $0$ and $\pi/2$.
 
\begin{figure}
 \centering
 \includegraphics[width=0.48\textwidth]{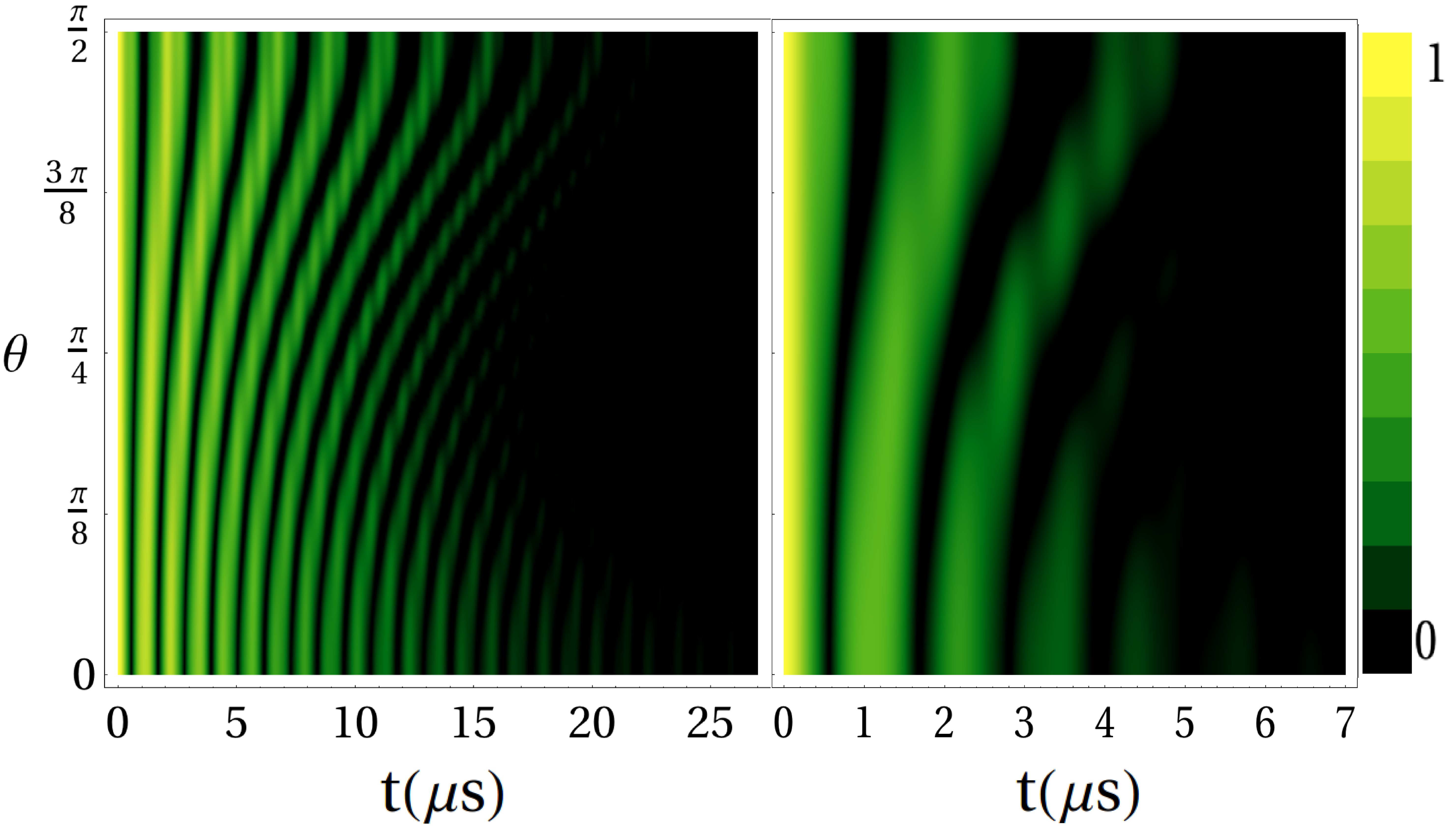}    
 \caption{(Color online) Entanglement density for different values of the rates $\gamma$ and $k$, varying the angles $\theta$ of inclination of the magnetic field of Earth ($47\mu T$). The entanglement was measured by the use of Concurrence. In the left panel, the rate values are $k=0.05MHz$, $\gamma=6k$, and in the right panel the values are $k=0.1MHz$, $\gamma=4k$. It's interesting to note that with smaller angles there are more entanglement sudden deaths and revivals, and that for angles around $\pi/4$ the decoherence time is shorter. The nuclear initial condition used in both cases was $\ket{\psi_{Nuclear}}=\big(\ket{0}+\ket{1} \big)/\sqrt{2}$. \label{fig:ConcurrenceDensity}}
 \end{figure} 
 
\subsection{The relevance of entanglement for angular sensitivity} 
Following the discussion in Ref. \cite{Cai}, the initial state in the RPM is not a perfect singlet (or triplet) state, so a natural way to test if the quantum correlations play a fundamental role in the working of the compass is to choose non-entangled initial conditions. In \cite{Cai} several random initial conditions were used, and the conclusion was that there is not a crucial dependence on the entanglement of the initial state, and that even RPs with initial separable states with only \textit{classical correlations} can produce an inclination sensitivity in the singlet yield.
We perturbed one thousand singlet initial conditions, some of them entangled, evolving under the density operator described in eq. \eqref{eq:MasterEquation} with the hyperfine tensor anisotropic, and with the decoherence processes turned off, i.e., $\gamma=0$}. 
To ensure real quantum states we distributed half of the sample using Bures measure \citep{Bures} defined as:
\begin{eqnarray}
 \rho_{S1}&=& \big( \mathbb{I}+U^{\dagger}\big)\tilde{\rho_S}\tilde{\rho_S}^{\dagger} \big( \mathbb{I}+U^{\dagger}\big),\\
 \rho_{S}&=& \frac{\rho_{S1}}{tr(\rho_{S1})}.
\end{eqnarray}
Here $U$ is a random unitary matrix distributed according to the Haar Measure \citep{Haar}, $\mathbb{I}$ is the identity matrix and $\tilde{\rho_{S}}$ is the perturbed singlet initial matrix. We proceeded in a similar way with the other half of the samples, distributing the perturbed state using the Hilbert-Schimdt measure:
\begin{eqnarray}
 \rho_{S}= \frac{\tilde{\rho_S}\tilde{\rho_S}^{\dagger}}{tr(\tilde{\rho_S}\tilde{\rho_S}^{\dagger})}.
\end{eqnarray}
With both renormalizations we ensure that our perturbed density matrix describes a true quantum state. The results of the simulations can be seen in Fig. \eqref{figSM:random}. As a first observation, we can see that the mechanism is robust. Even if some of the yields had no resemblance with the unperturbed system (black line in the figure), the mean gave the same amount of yield for each inclination of the field. This implies that the amount of entanglement is not a necessary condition to obtain angular sensitivity, given that none of the random initial conditions were in its maximum of entanglement and some of them were not entangled at all.
The results show that the amount of entanglement is not a fundamental factor for the sensitivity in the change of the angle of the applied magnetic field. A more specific example is presented in Fig. \eqref{fig:IsotropicSensibility} (red curve), with the initial state:
\begin{equation}\label{eq:initialstate}
\rho_{0}=\frac{1}{2}\big(\ket{\alpha\beta}\bra{\alpha\beta} + \ket{\beta\alpha}\bra{\beta\alpha}\big);
\end{equation}
this state gives an appreciable change in the yields (and therefore allow sensitivity) for different angles $\theta$.

\begin{figure}[ht!]
 \centering
 \includegraphics[width=0.44\textwidth]{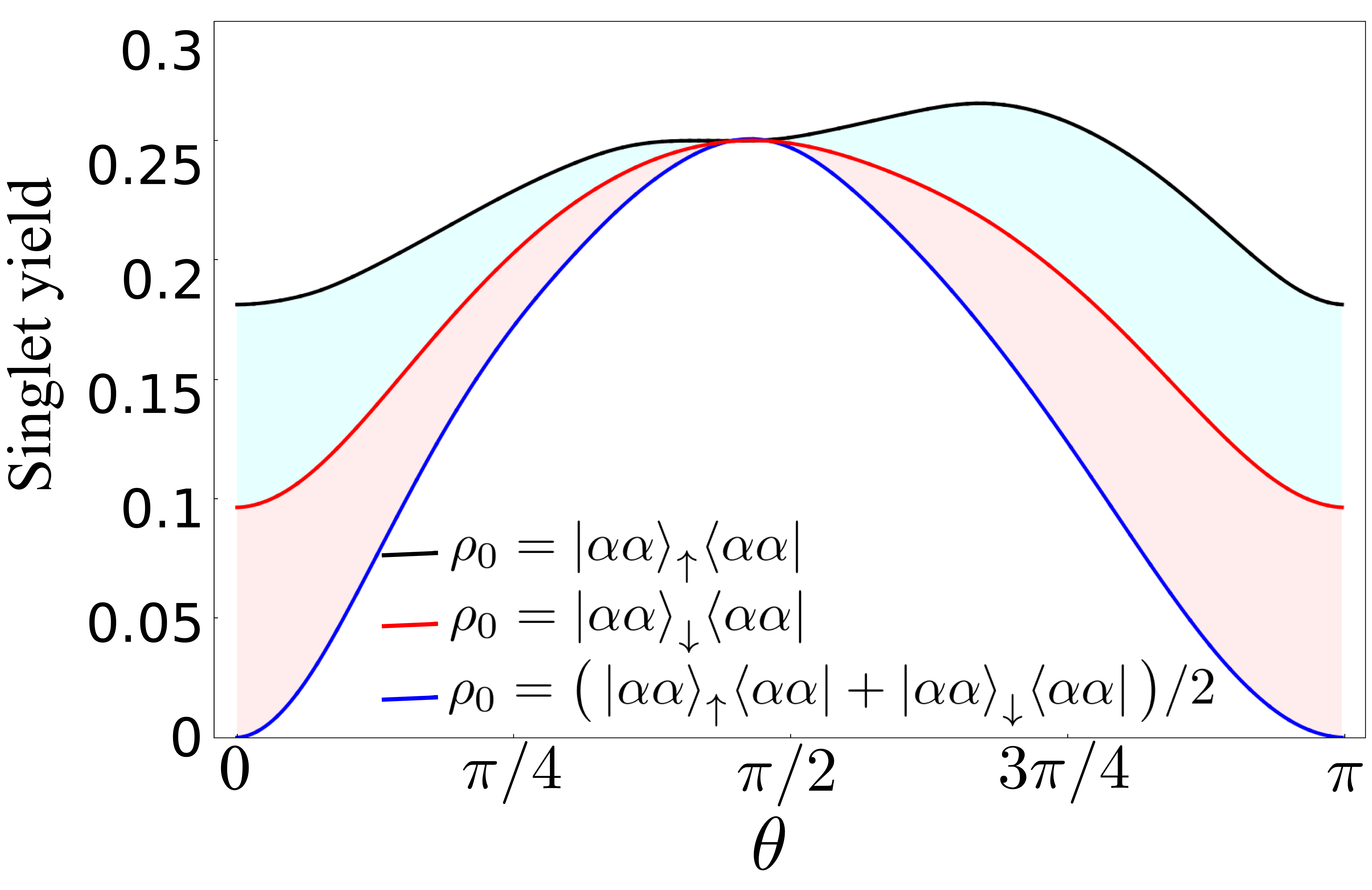}    
 \caption{(Color online) Angle sensitivity for non singlet or triplet initial conditions, specifically $\rho_{0}=\ket{\alpha\alpha}\bra{\alpha\alpha}$ with isotropic hyperfine tensors. Black (top curve): nuclear spin initially set to $\ket{\uparrow}$. Blue (lower curve): nuclear spin initially set to a mixed state $\big(\ket{\uparrow}+\ket{\downarrow}\big)/\sqrt{2}$. Red (middle curve): nuclear spin initially set to $\ket{\downarrow}$.\label{fig:Nuclearspin}}
\end{figure} 
This result implies that with an unentangled initial condition there is still a fully functional compass; in other words, the initial condition can be a source of anisotropy: if it is not an entangled one the system is still sensitive to changes in the inclination of the field. {This observation about non-entangled initial conditions was first made by Hogben \textit{et al.} \cite{Hore2012}}. In other words, in the absence of an explicit anisotropy in the hyperfine tensor or in the $g$ electronic factor, the sensitivity depends on the {\textit{inhomogeneity} of the populations in the density matrix}. As an example consider the state described in the blue curve in Figure \ref{fig:Nuclearspin}, i.e. a state with {only diagonal terms in its density matrix, evolving under an isotropic hyperfine coupling}: Its yield distribution shows a high variation with the angle. Moreover, an initial state like $\rho_{0}=\ket{\alpha\alpha}\bra{\alpha\alpha}$, which 
produces a lower but still appreciable angle sensitivity, generates a different distribution for the singlet yield depending on the nuclear spin state, {exemplifying the reach and relevance of this source of anisotropy}. {Surprisingly, after several numerical tests, only maximally entangled states (Bell states) will not generate any kind of sensitivity when the hyperfine tensor is isotropic}.

\subsection{Environmental induced anisotropy}
\begin{figure}[ht!]
 \centering
 \includegraphics[width=0.44\textwidth]{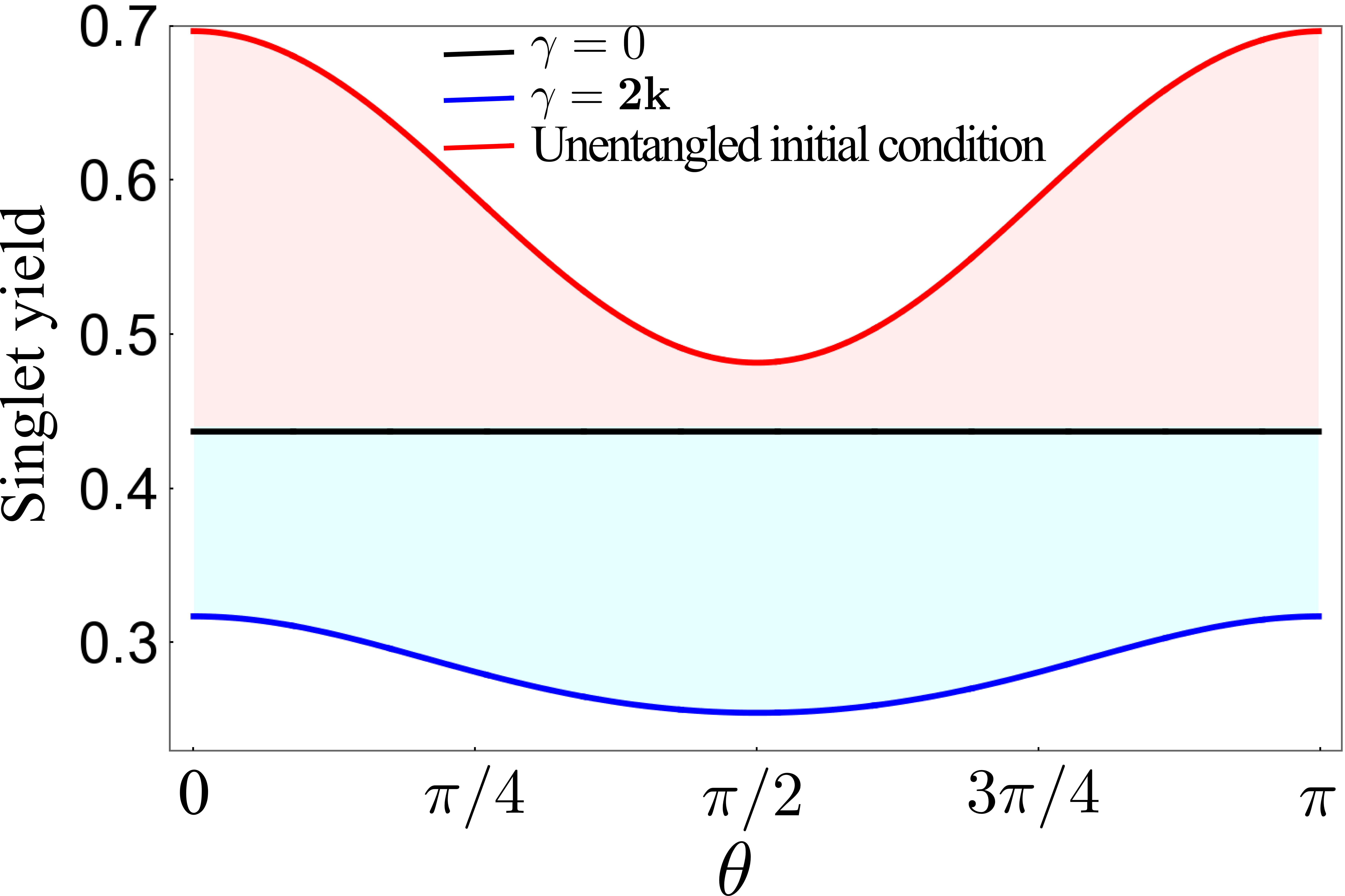}    
 \caption{(Color online) Angle sensitivity of isotropic Hamiltonians with $k=0.01MHz$. Black (middle curve): evolution of the singlet yield for different angles with an entangled initial condition. Blue (lower curve): evolution of the singlet yield for different angles with an entangled initial condition and $\gamma=2k$. Red (top curve): Evolution of the singlet yield for different angles with the unentangled initial condition {for both the electronic and nuclear spins $\rho_{0}=\frac{1}{4}\big(\ket{\alpha\beta}\bra{\alpha\beta} + \ket{\beta\alpha}\bra{\beta\alpha}\big) \otimes \big(\ket{\uparrow}\bra{\uparrow}+\ket{\downarrow}\bra{\downarrow}\big)$, and with $\gamma=0$}.\label{fig:IsotropicSensibility}}
 \end{figure}

If the Hamiltonian is isotropic, with singlet or triplet initial conditions, the expected behavior is an absence of change in the production rates of the yields, and there is going to be sensitivity only if there is a decoherence process present. This can be {understood} as another class of anisotropy induced by the environment, which chooses a preferred direction for the system through the dissipation. This can open the search of a suitable {chemical species} responsible for the RP creation, because the molecule does not need to present anisotropic hyperfine or Zeeman interactions, and the degree of entanglement is not going to be crucial. The only requirement for the correct functionality of the compass is then the decoherence, which is a must in such an open system. Both cases, anisotropy induced by the environment and by the initial conditions, can be seen in Figure \ref{fig:IsotropicSensibility}.
 
\section{Discussion}\label{disc}
Our findings go beyond establishing the role of the entanglement in the compass. We found that any kind of correlation, quantum or classical, is sufficient for the RPM if the molecule is anisotropic, and that an isotropic molecule can have sensitivity for variations in the field \textit{even} if it has classically correlated, separable initial conditions. One example of such a state is given by {eq. \eqref{eq:initialstate}}. Furthermore, even if they lack any kind of correlation but are \textit{anisotropic}, {giving an unbalanced weight to some populations over others}, like the states used in Figure \ref{fig:Nuclearspin}, we can also expect a working compass. A careful analysis has to be made in order to transparently identify the characteristics of the initial conditions that lead to a dependence of the singlet yield with the inclination angle of the field.

To avoid explicit environmental anisotropies, we used in the calculations only pure dephasing noises, avoiding the amplitude damping and thus changes in the energy levels of the system forced externally.

As discussed in \cite{Boxer1,Boxer2}, the anisotropy of a molecule can be averaged away if it has significant diffusive motion, or even rotations. Our findings of unexpected sources of anisotropy relax this immobility requirement. This is highly positive for the model given the wild conditions involving actual bio-photochemical processes. Along with this, isotropic molecules seem to be more robust to environmental effects; in the presence of a rf field the control experimental data\cite{Ritz2, Ritz1} shows that the compass will no longer work. However, if there is a strong enough dissipation the compass will work normally\cite{Gauger}. An isotropic molecule can handle higher noise magnitudes than an anisotropic one without jeopardizing control experimental data.

\begin{acknowledgments}
We thank M. Plenio and P. Hore for important comments and helpful discussion on our model.
This work is supported by the Brazilian funding agencies CNPq 
and FAPESP through the Instituto Nacional de Ci{\^e}ncia e 
Tecnologia - Informa\c{c}{\~a}o Qu{\^a}ntica (INCT-IQ). MCO wish to tank the hospitality of the Institute for Quantum Information Science at the University of Calgary, where part of this work was developed.
\end{acknowledgments}

\bibliographystyle{apsrev4-1}
\bibliography{AvianCompass}

\end{document}